\documentclass[%
 aip,
 jmp,%
 amsmath,amssymb,
preprint,%
]{revtex4-2}

\usepackage{graphicx}
\usepackage{dcolumn}
\usepackage{bm}
\usepackage{soul}
\usepackage{xcolor}

\begin{document}

\preprint{AIP/123-QED}

\title{Low barrier ZrO$_x$-based Josephson junctions}

\author{Jaehong Choi}
 \email[]{jc3452@cornell.edu}
\affiliation{ 
School of Applied and Engineering Physics, Cornell University, Ithaca, NY, 14850 
}

\author{Maciej Olszewski}%
\affiliation{ 
Department of Physics, Cornell University, Ithaca, NY, 14850
}

\author{Luojia Zhang}%
\affiliation{ 
Department of Physics, Cornell University, Ithaca, NY, 14850
}
\hfill

\author{Zhaslan Baraissov}
\affiliation{
School of Applied and Engineering Physics, Cornell University, Ithaca, NY, 14850
}

\author{Tathagata Banerjee}
\affiliation{
School of Applied and Engineering Physics, Cornell University, Ithaca, NY, 14850
}

\author{Kushagra Aggarwal}%
\affiliation{
School of Applied and Engineering Physics, Cornell University, Ithaca, NY, 14850
}

\author{Sarvesh Chaudhari}
\affiliation{
Department of Physics, Cornell University, Ithaca, NY, 14850
}

\author{Tom\'as A. Arias}
\affiliation{
Department of Physics, Cornell University, Ithaca, NY, 14850
}

\author{David A. Muller}
\affiliation{
School of Applied and Engineering Physics, Cornell University, Ithaca, NY, 14850
}

\author{Valla Fatemi}
\affiliation{
School of Applied and Engineering Physics, Cornell University, Ithaca, NY, 14850
}

\author{Gregory D. Fuchs}
 \email[]{gdf9@cornell.edu}
\affiliation{
School of Applied and Engineering Physics, Cornell University, Ithaca, NY, 14850
}

\date{\today}

\begin{abstract}
The Josephson junction is a crucial element in superconducting devices, and niobium is a promising candidate for the superconducting material due to its large energy gap relative to aluminum. AlO$_x$ has long been regarded as the highest quality oxide tunnel barrier and is often used in niobium-based junctions. Here we propose ZrO$_x$ as an alternative tunnel barrier material for Nb electrodes. We theoretically estimate that zirconium oxide has excellent oxygen retention properties and experimentally verify that there is no significant oxygen diffusion leading to NbO$_x$ formation in the adjacent Nb electrode. We develop a top-down, subtractive fabrication process for Nb/Zr-ZrO$_x$/Nb Josephson junctions, which enables scalability and large-scale production of superconducting electronics. Using cross sectional scanning transmission electron microscopy, we experimentally find that depending on the Zr thickness, ZrO$_x$ tunnel barriers can be fully crystalline with chemically abrupt interfaces with niobium. Further analysis using electron energy loss spectroscopy reveals that ZrO$_x$ corresponds to tetragonal ZrO$_2$. Room temperature characterization of fabricated junctions using Simmons’ model shows that ZrO$_2$ exhibits a low tunnel barrier height, which is promising in merged-element transmon applications. Low temperature transport measurements reveal sub-gap structure, while the low-voltage sub-gap resistance remains in the megaohm range. 
\end{abstract}

\maketitle

%

\section{\label{sec:level1} Introduction}

Josephson junctions (JJs) form the essential nonlinear circuit element in superconducting qubits, which are a leading platform for realizing scalable, large-scale quantum computation.~\cite{Clarke2008,Krantz2019,Siddiqi2021,Kjaergaard2020} From a materials standpoint, aluminum is the standard superconducting electrode material in JJs due to the reproducibility of its thin-film deposition, well-established chemistry, and the robust Josephson effect in an Al/AlO$_x$/Al structures.  The AlO$_x$ tunnel barrier is formed by thermal oxidation of the Al electrode.\cite{Biznarova2024} On the other hand, aluminum's low critical temperature makes it susceptible to quasiparticle poisoning,\cite{Leppakangas2012} thus limiting further improvements in qubit coherence\cite{Martinis2009, Catelani2011, Catelani2014, Serniak2018, Connolly2024, Meservey1971}, operating frequency, and operating temperature. In considering alternatives, niobium has emerged as promising because it can be sputtered in high quality and it has a wide superconducting gap relative to aluminum.\cite{Finnemore1966} 

In a Nb-based JJ material platform, AlO$_x$ has also been adopted as a tunnel barrier of choice.\cite{Anferov2024, Anferov20242} However, AlO$_x$ and Nb as a barrier electrode material system face some challenges, which has prevented the widespread adoption of this platform for quantum technologies. First, the chemical and thermal instability of AlO$_x$ relative to NbO$_x$ can lead to oxygen diffusion from the barrier layer into the surrounding electrodes.\cite{Tan2005, IEEE2021, Tolpygo2010, Verjauw2021} Additionally, there have been studies reporting that NbO$_x$ in base layers is a major source of microwave loss in resonators and cavities,\cite{Burnett2016,Romanenko2017, Romanenko2020, Niepce2020} which raises the question as to whether NbO$_x$ in the JJ itself could be problematic. For example, recent studies have shown that NbO$_x$ likely contributes to two-level system (TLS) loss, resulting in energy relaxation in a transmon qubit.\cite{Premkumar2021, Bal2024} Meanwhile, encapsulating a Nb electrode with Ta decreases the relaxation rate.\cite{Bal2024}

Several studies have investigated methods to mitigate oxygen diffusion from AlO$_x$ in Nb-based JJs. Anferov et al.~used an Al layer as an oxygen diffusion barrier to suppress the NbO$_x$ formation and demonstrated an average qubit quality factor\cite{Anferov2024} above 10$^5$ and qubit operation temperature up to 200 mK with relaxation and dephasing times of approximately 1~$\mu$s.\cite{Anferov20242} There is also a concern regarding the amorphous nature of AlO$_x$, as it has been speculated that the coherence of Josephson qubits with an AlO$_x$ tunnel barrier is disrupted by spectral splittings indicating TLS in AlO$_x$.\cite{Phillips1987, Martinis2005,Shnirman2005,Ku2005,Faoro2006} To address this concern, Oh et al. studied superconducting qubits fabricated with an epitaxially-grown crystalline Al$_2$O$_3$ tunnel barrier and showed that they exhibit fewer TLSs compared to qubits with amorphous AlO$_x$.\cite{Oh2006} Also, Kim et al. explored crystalline epitaxial AlN tunnel barriers in all-nitride NbN/AlN/NbN JJ and showed their high chemical stability against oxidation.\cite{Kim2021}

In this work, we show that ZrO$_x$ is a promising alternative tunnel barrier material for Nb-based JJs. While a few studies have explored tunnel junctions with ZrO$_x$ barriers,\cite{Wang2001, Nevirkovets2006} systematic investigations into its application in superconducting quantum devices remain unexplored. Cross-sectional scanning transmission electron microscopy (STEM) indicates that the ZrO$_x$ barrier can be crystalline ZrO$_2$ with a tetragonal phase. A crystalline tunnel barrier may be advantageous for qubit stability and it may help to limit TLS-based decoherence from the JJ itself. In addition, instead of adopting the traditional Niemeyer-Dolan bridge structure\cite{Niemeyer1976, Dolan1977} or Manhattan structure,\cite{Potts2001, Lecocq2011} we implement a top-down, subtractive fabrication process for JJs from continuous Nb/Zr+oxidation/Nb films.  This approach enables improved process control over the tunnel barrier formation relative to traditional additive fabrication. In our process, we define the bottom electrode and tunnel barrier using masked etching, which provides precise control over junction geometry and is compatible with large-scale fabrication using semiconductor techniques. We demonstrate that ZrO$_2$ possesses excellent oxygen retention, low barrier height, CMOS compatibility, and high subgap resistance that make it a promising candidate for quantum device applications, including merged-element transmons.\cite{Zhao2020, Mamin2021}

\section{\label{sec:level1} Oxygen retention of zirconium oxide}

\begin{figure}[h!]
\includegraphics[width=\textwidth]{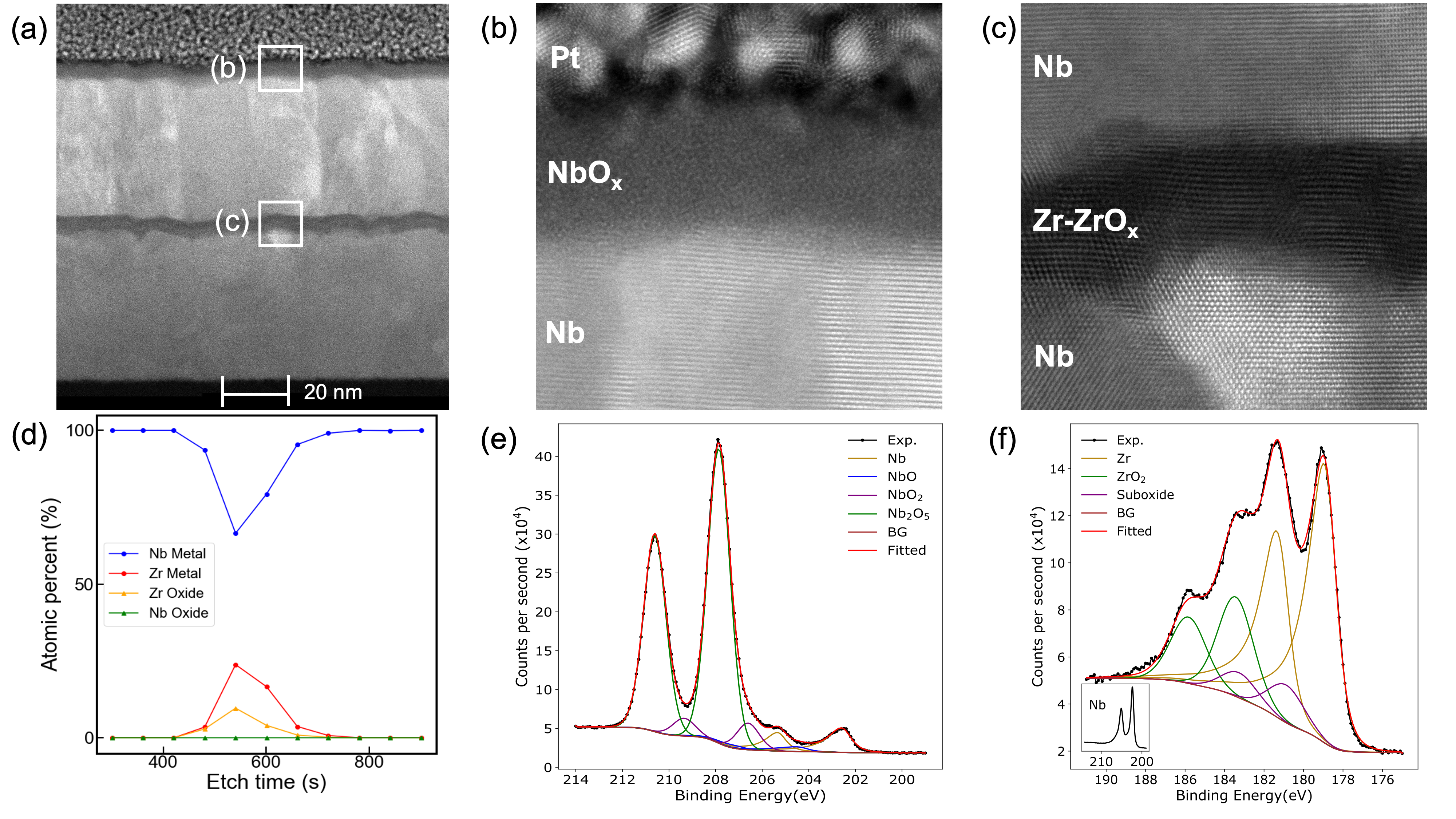}
\caption{\label{fig:wide} (a)~HAADF-STEM image of the Nb/ZrO$_x$-Zr/Nb films with a nominal Zr thickness of 5~nm. Magnified views of (b)~the top region, showing Pt/NbO$_x$/Nb where a thin NbO$_x$ layer forms on the top Nb electrode, and (c)~the top Nb/Zr–ZrO$_x$/bottom Nb region where no NbO$_x$ is observed between the top Nb electrode and the ZrO$_x$ tunnel barrier.(d)~Relative atomic percentages of Nb, NbO$_x$, Zr, and ZrO$_x$ in the quadri-layer based on XPS analysis. No NbO$_x$ signal was detected. (e)~Nb 3d core level XPS scan of the top Nb electrode region (pre-etch) and (f)~the Zr 3d core level XPS scan of the Zr–ZrO$_x$ barrier region at the point of maximum Zr signal (after 540s of etching). The inset shows the XPS signal from Nb, indicating the absence of oxide.}
\end{figure}

Native AlO$_x$ is a high quality tunnel barrier with low leakage and low dielectric loss. However, it has a less negative enthalpy of formation ($\Delta H_{f}$=–1675 KJ mol$^{-1}$) than that of niobium oxide~($\Delta H_{f}$=–1903 KJ mol$^{-1}$), which suggests that oxygen may thermodynamically favors the bonding with Nb to form NbO$_x$ rather than the formation of AlO$_x$. The relative enthalpies of formation provide a thermodynamic perspective on the tendency of oxygen migration. Such oxygen instability of AlO$_x$ could be problematic in Nb-based JJs because NbO$_x$ has been speculated to contribute to energy relaxation in transmon qubits.\cite{Premkumar2021, Bal2024}

To address the inherent thermodynamic instability in Nb/AlO$_x$/Nb interfaces, we explore alternative tunnel barrier materials for Nb-based JJs. Specifically, we look for tunnel barrier materials that satisfy the following criteria, quantified by first-principles energy calculations: 1)~higher vacancy formation energy in the oxide compared to its metallic form, which indicates that oxygen prefers to remain in the desired oxide, rather than diffuse into the metal. To evaluate this, we calculate $\Delta$E$_{Mo \rightarrow M}$, the energy required for an oxygen atom to move from native oxide form (Mo) to metallic form (M), leaving behind a vacancy. 2)~More stable oxygen interstitials in the barrier material than in Nb, so that even if oxygen is introduced into the tunnel barrier, it is less likely to migrate into the Nb and form NbO$_x$. We quantify this with $\Delta$E$_{M \rightarrow Nb}$, the energy required for an oxygen interstitial to diffuse from M into the Nb electrode. Finally, 3)~higher energy required to remove an oxygen interstitial from the tunnel barrier than to fill in an oxygen vacancy in NbO$_x$. This corresponds to $\Delta$E$_{M \rightarrow NbO_x}$, the energy needed for oxygen from M to complete NbO$_x$. Materials with all positive $\Delta$E values are preferred, as this indicates that the processes leading to NbO$_x$ formation are energetically unfavorable.

We calculate these energies using a density functional study based on plane wave bases and periodic boundary conditions within the generalized gradient approximation (see Appendix A for more details). This theoretical work is in preparation and more details will be available in the pre-print soon. Among the metals we examined, Zr is the only material that gives positive values for all three energy terms: $\Delta$E$_{Mo \rightarrow M}$=0.19~eV, $\Delta$E$_{M \rightarrow Nb}$=2.56~eV, and $\Delta$E$_{M \rightarrow NbO_x}$=0.17~eV. This suggests that ZrO$_x$ has strong oxygen retention properties when in contact with metallic Nb.

To test these ideas, we first sputter 60~nm of Nb at 2~mTorr and with a 70~sccm flow of ultra high-purity Ar gas. Following the Nb deposition, we deposit Zr with the same process gas parameters. Upon completion of the Zr deposition, the chamber is purged and subsequently flooded with a 15\% oxygen in argon mixture with a 100~sccm flow rate for 16 minutes (3700~Torr~s). After continuous gas flow, the pressure reaches approximately 4.2~Torr, at which point the oxidation gas is purged. The chamber is then pumped for around 2 to 3 hours to remove excess oxidation gas. Before depositing the top Nb layer, the Nb target is pre-sputtered for 5 minutes. We deposit the top Nb electrode to a thickness of 60~nm using the same parameters as the bottom Nb electrode. All these processes are done at room temperature. 

We explore Zr thicknesses ranging from 1.5~nm to 5~nm while maintaining identical pressure and oxidation time, with the idea that for there are Zr thicknesses at which that the Zr will be fully oxidized.
While we primarily focus on junctions fabricated from junction layers with a 5~nm-thick Zr layer because they show the least sample-to-sample variation, and illustrate nearly all key features of these devices in transport with reliability, we also present some transport and electron microscopy results for junctions with thinner Zr junctions where the Zr is fully oxidized into ZrO$_2$ as discussed below. 
We verify the oxygen retention of ZrO$_x$ using STEM and X-ray photoelectron spectroscopy (XPS). Fig.~1a-c show dark-field STEM cross-sectional images of the Nb/Zr~(5~nm)/O$_2$ 3700~Torr~s/Nb. We observe a thin layer NbO$_x$ has formed on the Nb top electrode due to exposure to ambient air (Fig.~1b), however, no NbO$_x$ is observed between the top Nb electrode and Zr-ZrO$_x$ layer in Fig.~1c. The slight color contrast at the edge of the top Nb film in Fig.~1c is due to a boundary plane in projection that is not parallel to the beam direction. In the magnified STEM images (Fig.~1b-c) the crystalline Nb electrode is clearly visible. In Fig.~1c, we observe that the Zr-ZrO$_x$ layer exhibits crystalline order with chemically sharp interfaces with the niobium electrodes. This suggests crystalline ZrO$_x$. 

Fig.~1d shows the relative atomic percentage of the Nb, Zr metals, and their respective oxides obtained from depth-dependent XPS after milling through the top oxide. The depth profile is made by Ar ion milling through the trilayer and taking XPS data every 60s. The relative atomic percentage is calculated by fitting the Nb 3d, O 1s, and Zr 3d core level scans at each etch layer. Fig.~1e shows the Nb 3d scan on the top of the sample pre-etch, indicating a highly oxidized surface. Fig.~1f shows the Zr 3d scan at the point with maximum Zr signal, after 540s of etching. The Zr oxide content is calculated as the summation of the Zr$^{4+}$ and the Zr suboxide components extracted from the fitting shown in Fig.~1f, though we note that Zr$^{4+}$ comprises the majority of the signal (see Appendix G for details). 
We observe no Nb oxides in the Nb spectrum throughout the depth profile, and the profile away from the ZrO$_x$ layer is identical to that of pure Nb films.

\begin{figure*}
\includegraphics[width=\textwidth]{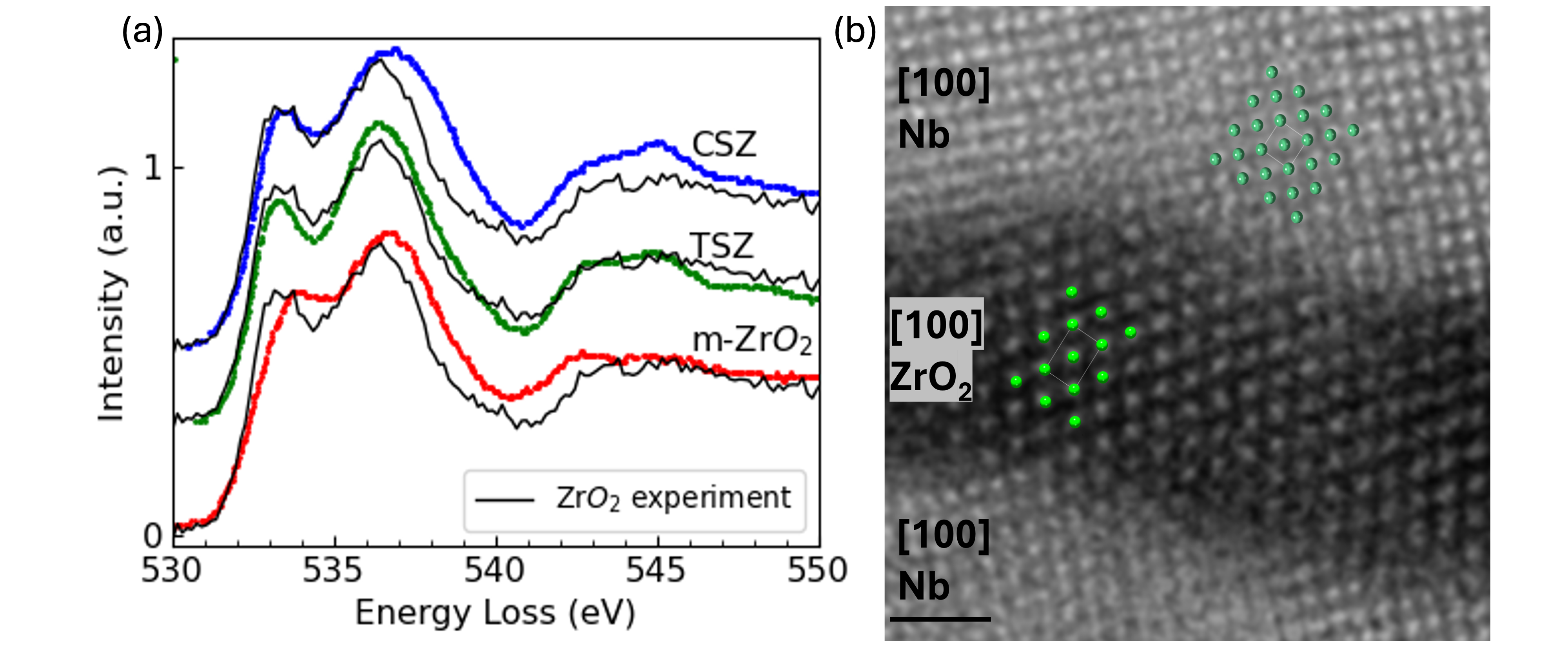}
\caption{\label{fig:wide} (a)~Comparison of the EELS signal to the spectra adopted from McComb~\cite{McComb1996}. Colors correspond to different crystal symmetries. Here, the tetragonal stabilized zirconia shows the best match to the ZrO$_2$, used in this experiment. (b)~High-resolution HAADF STEM image of the tunneling barrier that was formed from 1.5 nm of Zr prior to oxidation.  Afterward, its width thickened to 2.1~nm. The top and bottom Nb electrodes are oriented at [100] zone axis. Zr atoms in this image are then overlaid with the Zr atoms in [100] ZrO$_2$ orientation. For reference, Nb atoms in electrodes are overlaid with its associated structure. The size of the scalebar in the left bottom of the image is 1~nm. }
\end{figure*}

To elucidate the crystal structure of the ZrO$_x$ tunnel barrier, its stoichiometry, and the interface between the oxide and the Nb electrodes, we employ further electron microscopy on a film that was grown with 1.5~nm of Zr and then oxidized. When identifying stoichiometry of ZrO$_x$, traditional phase identification methods such as nano-beam electron diffraction (NBED) and STEM imaging proved insufficient. The limitations arise from the overlapping grain structures, small crystallite sizes, and the very subtle structural distinctions between the monoclinic, tetragonal, and cubic phases. As pointed out by McComb,\cite{McComb1996} differentiating t-ZrO$_2$ and c-ZrO$_2$ in the presence of m-ZrO$_2$ is particularly challenging due to their closely related crystallographic parameters and nearly identical diffraction profiles. Moreover, real-space methods, such as HAADF and iDPC, struggle to clearly resolve the fine atomic-scale distortions and local symmetry variations that differentiate these phases, mainly due to inability of most imaging methods to clearly resolve the oxygen atoms. 

To overcome these limitations, we employ EELS, focusing on the oxygen K-edge, which provides distinct spectroscopic “fingerprints” for each polymorph.\cite{McComb1996} The near-edge fine structure (ELNES) reflects the unoccupied density of states, which is strongly influenced by the symmetry and coordination of the oxygen and zirconium atoms. For example, at the oxygen K-edge, the positions and shapes of the peaks are dictated by hybridization between oxygen 2p and zirconium 4d states, which are split by the crystal field in ways that differ across the monoclinic, tetragonal, and cubic phases. As the concentration of ZrO$_7$ polyhedra and the amount of oxygen vacancies decrease from cubic to monoclinic ZrO$_2$, McComb showed that the energy separation between the first two peaks in the oxygen K-edge spectrum also decreases systematically from cubic to monoclinic ZrO$_2$, and that variations in peak width and separation can distinguish these phases.\cite{McComb1996} Since EELS spectra are sensitive to local symmetry, the effects of oxygen vacancies, distortions, and dopants result in broadening or shifts in peak position. This makes EELS valuable when structural differences manifest more strongly in the local electronic environment than in long-range order, as is the case with stabilized ZrO$_2$ polymorphs and nanocrystalline samples.

To determine the oxidation state of Zr, we first examine the onset of the oxygen K-edge, which begins at approximately 532.5 eV. This onset is consistent with that reported for fully oxidized ZrO$_2$ phases, effectively ruling out substoichiometric phases such as Zr$_3$O, Zr$_2$O, or ZrO, whose O-K edge onsets appear at higher energies. Furthermore, to identify the specific ZrO$_2$ phase, we overlay reference spectra for the three ZrO$_2$ phases in Fig.~2a with the experimental EELS spectrum from the green, fully oxidized region in Fig.~4g, based on McComb’s work. The experimental spectrum aligns closely with the tetragonal stabilized zirconia (TSZ) signature. Notably, the energy splitting between the two leading peaks matches the $\sim$3 eV separation characteristic of TSZ. In contrast, the cubic stabilized zirconia (CSZ) shows a larger peak separation, and the monoclinic phase (m-ZrO$_2$) exhibits a pronounced low-energy shoulder that is absent in our data. The identification of the TSZ phase also provides insight into the concentration of oxygen vacancies. Prior studies\cite{Steele1974, Li1993} have shown that TSZ is stabilized at relatively low concentrations of vacancy-inducing dopants such as MgO, CaO, or Y$_2$O$_3$, while higher dopant levels favor the formation of CSZ. This suggests that the TSZ phase in our sample corresponds to an oxygen vacancy concentration of below ~20\%, above which the structural transition to the CSZ phase is typically observed. Detailed information on EELS acquisition and segmentation can be found in Appendix C and Fig.~7.

Figure~2b shows a high-resolution HAADF image of the Nb/ZrO$_2$/Nb interface. The nominal thickness of Zr in this film was 1.5~nm before oxidation and it thickened to 2.1~nm by oxidation. Both Nb grains, located above and below the oxide barrier, are oriented along the [100] zone axis. Within the ZrO$_2$ region, although the lighter oxygen atoms are not resolved in HAADF imaging, the positions of the heavier Zr atoms can be indexed to the [100] orientation of tetragonal ZrO$_2$. This structural assignment was confirmed using the crystallographic model for mp-2574 retrieved from the Materials Project (database version v2025.04.10). The interface between Nb and ZrO$_2$ appears atomically sharp and coherent, with no observable interfacial disorder. This observation suggests that epitaxial growth of ZrO$_2$ on Nb under the right conditions could yield a structurally perfect single-crystal oxide barrier. Such a defect-free interface has significant implications for applications that require high-quality tunnel barriers, where structural coherence and stoichiometric control are essential for improving qubit performance.

\section{\label{sec:level1} Josephson Junction Fabrication}

\begin{figure*}
\includegraphics[width=\textwidth]{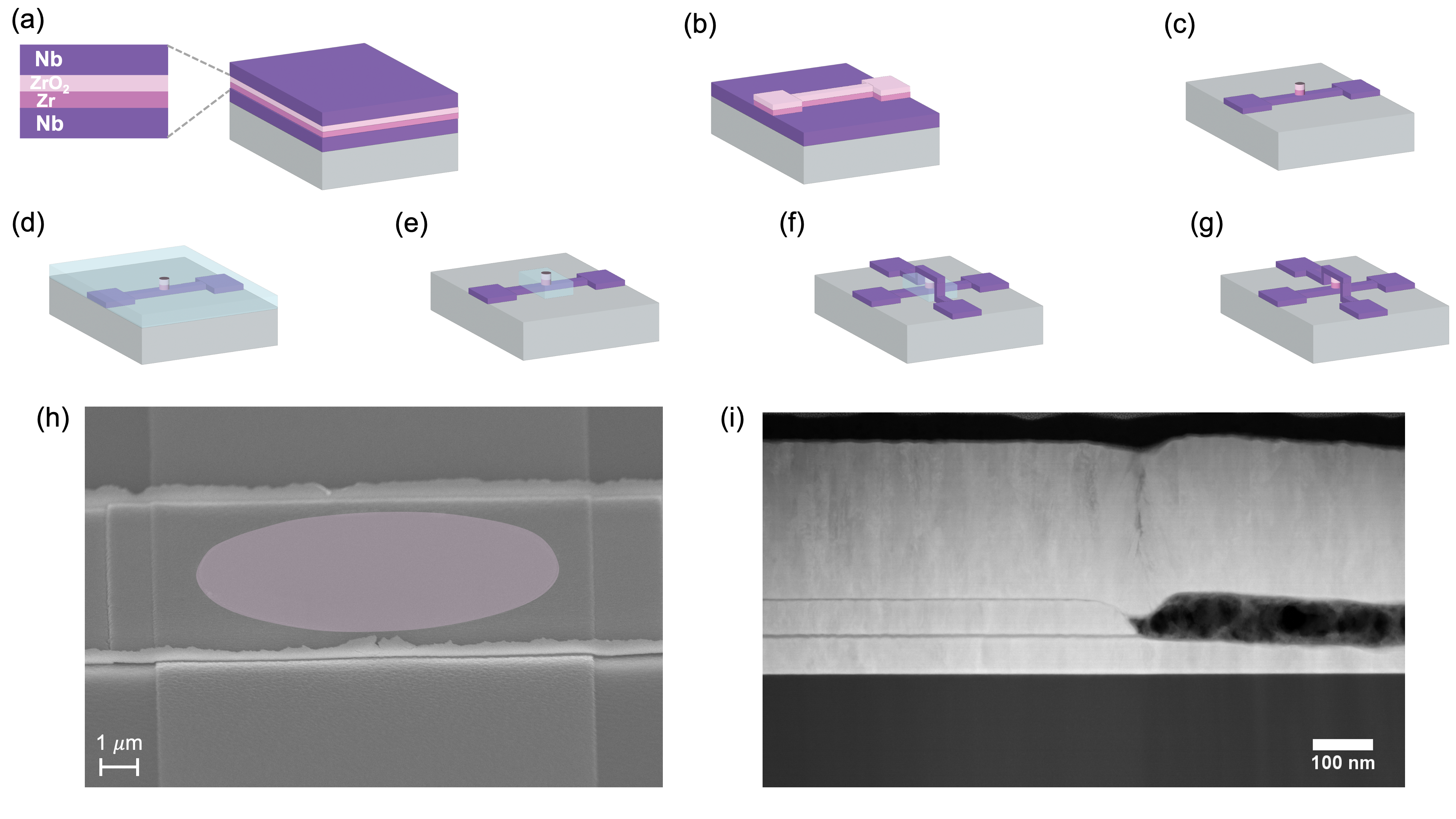}
\caption{\label{fig:wide} JJ fabrication process and SEM/STEM images of the junction (a)~Nb/Zr~(5~nm)/O$_2$ 3700~Torr~s/Nb quadri-layer sputtering with in-situ oxidation. (b)~We define the bottom electrode by performing ICP-RIE etching using Cl$_2$,BCl$_3$,and Ar plasma to partially etch the trilayer. (c)~A second ICP-RIE etching to define the cylindrical tunnel barrier. (d)~SiO$_2$ spacer evaporation right after tunnel barrier etching without resist removal. (e)~After lift-off of the SiO$_2$ spacer, BOE is used to etch the SiO$_2$ spacer layer. (f)~The top Nb electrode is sputtered after resist removal. (g)~We remove the SiO$_2$ spacer using vapor HF etching. (h)~False-colored SEM image of the resultant junction. The pink region indicates the circular junction. (i)~STEM cross-section edge image of the junction before SiO$_2$ scaffold removal.
}
\end{figure*}

To study the potential of ZrO$_2$ as a tunnel barrier, we develop a top-down, subtractive fabrication process that starts with a continuous film of Nb/Zr/O$_2$ 3700~Torr~s/Nb. The bottom electrode and tunnel barrier are defined by etching rather than through shadow-evaporation. This approach enables precise and consistent control over feature size and shape, making it compatible with current semiconductor manufacturing techniques and suitable for large-scale fabrication of superconducting quantum processors. We also implement an airbridge top electrode, similar to several recent studies,\cite{Anferov2024,Anferov20242, Gronberg2017,Ke2025} to reduce the need for an extra dielectric scaffold structure, which can be a potential source of loss.\cite{Ke2025} Additionally, we shape the tunnel barrier cylindrically to allow sidewall accesses from all angles for isotropic characterization. 

Figures~3a-g outline the JJ fabrication process. First, we pattern the bottom electrode using photolithography and partially etch the Nb/Zr-ZrO$_2$/Nb trilayer using Cl$_2$, BCl$_3$, and Ar plasma (Fig.~3b). After this step, we remove the photoresist and repeat the photolithography and etch processes to define the cylindrical tunnel barrier (Fig.~3c). Without removing the photoresist, we load the sample into the evaporator and deposit an SiO$_2$ spacer layer to electrically isolate the bottom and top electrodes (Fig.~3d). Lift-off removes the SiO$_2$ above the cylindrical tunnel barrier. We use photolithography to pattern a square-shaped SiO$_2$ scaffold with a 3~$\mu$m overhang extending from the tunnel barrier on all four sides. We etch the SiO$_2$ using a buffered oxide etch (BOE) solution (Fig.~3e) everywhere except the square area covered by the photoresist. After resist removal, a 300~nm thick top Nb electrode is deposited using a lift-off process (Fig.~3f), followed by vapor hydrofluoric acid (HF) etching to remove the SiO$_2$ scaffold (Fig.~3g). The scanning electron microscopy (SEM) (Fig.~3h) and STEM cross-section edge image of the resultant JJ (Fig.~3i) show that the free-standing top and bottom electrodes are separated by a small gap, which confirms that the top electrode does not collapse after its supports is removed by vapor HF etching. 

\section{\label{sec:level1} Room temperature JJ characterization and tunneling parameters}

\begin{figure*}
\includegraphics[width=\textwidth]{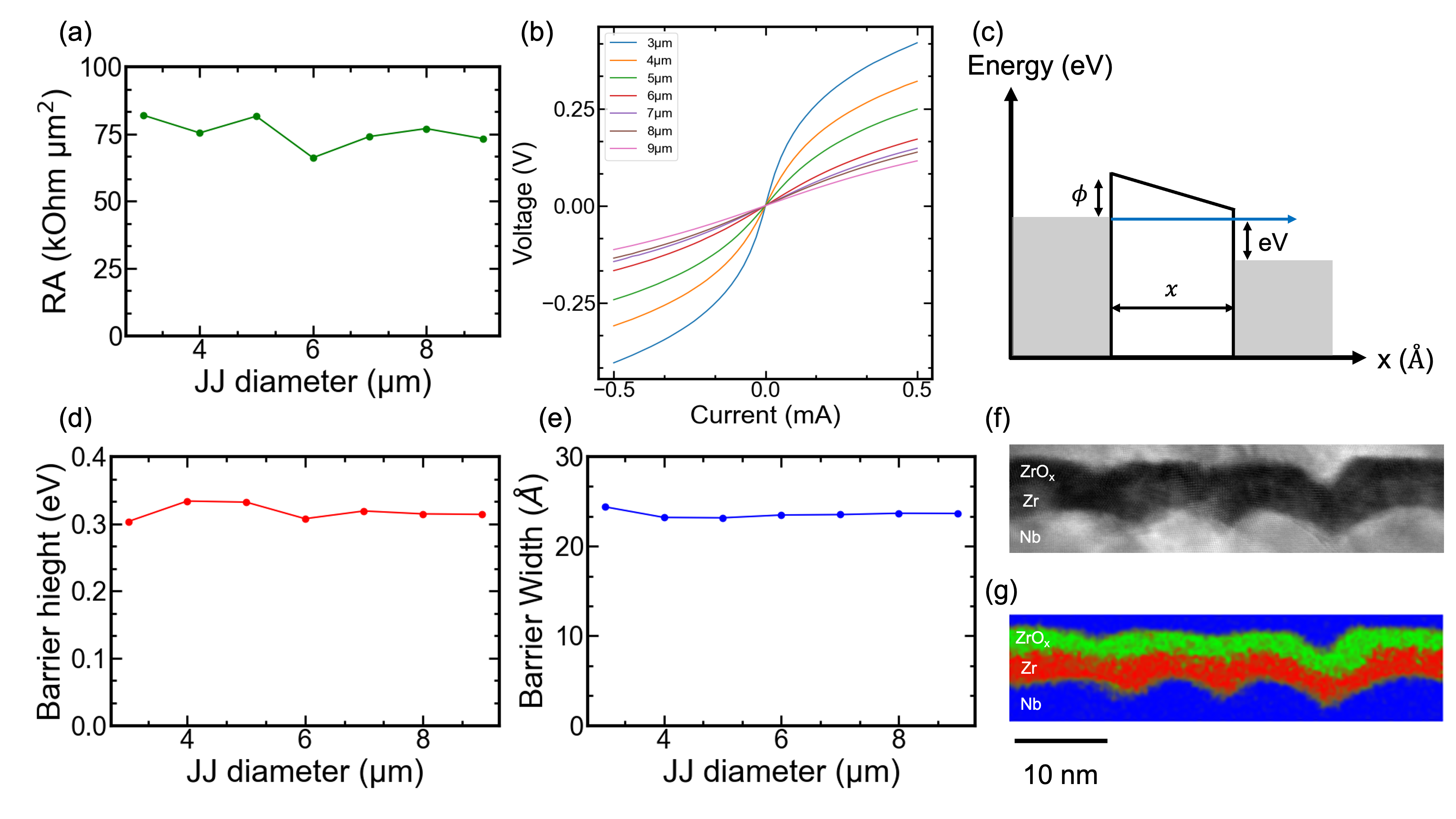}
\caption{\label{fig:wide} Room temperature I-V characteristics and tunneling parameters of junctions grown with 5~nm of Zr. (a)~RA products for junctions with varying diameters, ranging from 2~$\mu$m to 9~$\mu$m. (b)~Current-voltage curves from junctions with different diameters. (c)~Energy diagram of of a metal-insulator-metal tunnel junction. The shaded area indicates filled electron state up to fermi-level, and when a small voltage is applied, the tunnel barrier is tilted and allows electrons to tunnel through the barrier and induces non-linear conduction. $\phi$ is tunnel barrier height and $x$ is tunnel barrier width. Extracted (d)~tunnel barrier height and (e)~width using Simmons' model. (f)~ADF-STEM image of the Nb/Zr-ZrO$_2$/Nb trilayer. (g)~STEM-EELS segmentation study of the trilayer. It reveals that 2.5~nm of 5~nm Zr was oxidized, which is consistent with the Simmons' model estimation of the tunnel barrier width. }
\end{figure*}

We can evaluate the junction quality by measuring the RA product of several junctions at room temperature. The RA product is defined as the product of the junction resistance R and junction sectional area A. Its invariance across devices indicates structural and electrical homogeneity of the tunnel barrier.\cite{Boyn2016} The RA value is ideally determined by material properties of the tunnel barrier and electrodes, and constant RA values as a function of junction diameter further implies the absence of resistive sidewall shunting caused by material redeposition during or after etching.  If sidewall shunting is present, the RA values will decrease for the smallest junctions. We examine the RA of junctions with diameters ranging from 3~$\mu$m to 9~$\mu$m and observe roughly constant RA values (Fig.~4a), which indicates a high-quality tunnel junction process. 

Next, we study the I-V characteristics of the junctions at room temperature. We conduct current-biased four-wire measurements and obtain nonlinear I-V curves from junctions formed from 5 nm of Zr across all junction diameters (Fig.~4b). These curves are characteristic of tunneling behavior. Notably, the current becomes nonlinear at low voltages, which indicates a low tunnel barrier height. To study the tunnel barrier characteristics of ZrO$_2$ in more detail, we employ Simmons' model, which describes the tunnel current density~(J) in the low voltage regime for a system in with two metal electrodes separated by a thin insulating film in the WKB approximation\cite{Simmons1963,Simmons19632}:        
\begin{equation}
    J = \beta (V+\gamma V^3)
\end{equation}
where $\beta (x,\phi) = \frac{3}{2} \frac{e^2 (2m\phi)^{1/2}}{h^{2}x} exp(-A \phi^{1/2})$, $\gamma(x,\phi) = \frac{(Ae)^2}{96\phi}-\frac{Ae^2}{32\phi^{3/2}}$, and $A=\frac{4 \pi (2m)^{1/2} x}{h}$, $m$ is the electron mass, $h$ is Planck's constant, $\phi$ is the tunnel barrier height, and $x$ is the tunnel barrier width. 

\begin{figure*}
\includegraphics[width=\textwidth]{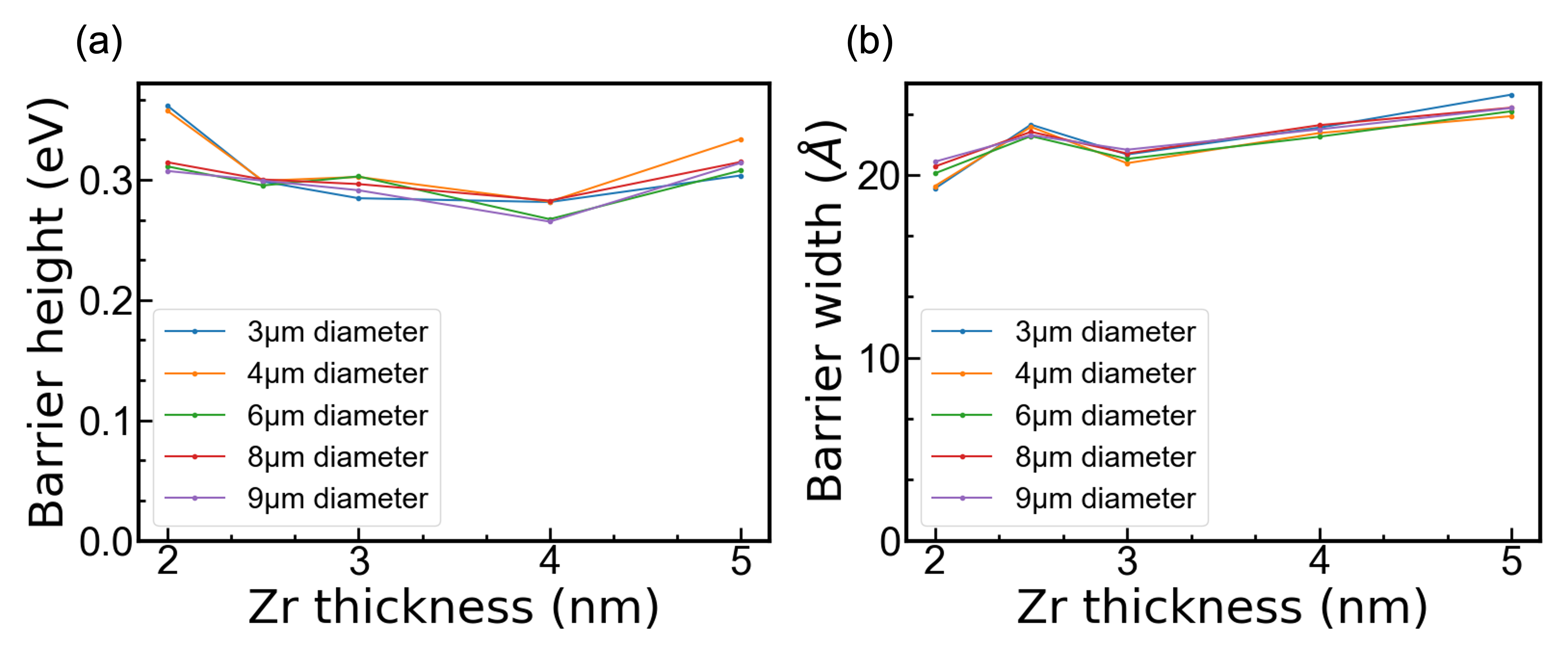}
\caption{\label{fig:wide} (a)~Simmons' model extracted barrier height and (b)~width of the junctions with varying Zr thicknesses and diameters.}
\end{figure*}

By fitting the J-V data with Equation~(1), we can extract tunneling parameters – tunnel barrier height~($\phi$) and tunnel barrier width~($x$) (Fig.~4c). We observe consistent tunnel barrier height and width across junctions with different diameters (Fig.~4d-e). This again implies a high quality fabrication process. Also, consistent tunneling parameters suggest the stability of the Simmons' model across different feature metrics. The Simmons' model estimates that the tunnel barrier height of ZrO$_2$ is 0.3~eV, as shown in Fig.~4d. This is an order-of-magnitude smaller than the tunnel barrier height of a Nb-based JJ with an AlO$_x$ tunnel barrier (see more details in the Appendix). 
In contrast, Simmons' model fits report an estimated a barrier height of 3~eV for a 1.2~nm thick AlO$_x$ in a Nb/AlO$_x$/Nb tunnel junction\cite{Holmqvist2008} and 1~eV for a 1~nm AlO$_x$ in an asymmetric Al/AlO$_x$/Nb junction.\cite{Jung2009} These all exhibited higher barrier heights despite having lower thicknesses.
Using alternative methods, a Nb/Al-AlO$_x$/Nb tunnel junction with an ultrathin AlO$_x$ layer (0.6–1.5~nm) are reported to have a barrier height of approximately 1.2~eV based on ballistic electron emission microscopy (BEEM) and a scanning tunneling spectroscopy.\cite{Rippard2002} This is important because the critical current decays exponentially with barrier thickness, and thinning the tunnel barrier to achieve higher critical currents for qubit operations pose a narrower process window, thus making it difficult to ensure uniformity, and control pinhole formation. 

A tunnel barrier with a low barrier height is particularly interesting in terms of merged-element transmon (MET) devices. In METs, the internal capacitance of JJs replaces an external shunt capacitor. This requires a careful design of JJs to simultaneously achieve a capacitance that is large enough to reduce the charge dispersion of energy while maintaining sufficient anharmonicity to isolate two energy levels relative to the other transitions. This balance can be achieved for junction areas that are lithographically easy to print by designing a junction with a low barrier height, allowing the use of a thick barrier to reduce the large critical current that would otherwise be induced by the large JJ dimensions needed for increased capacitance. Additionally, JJs with a thick barrier can reduce the variation of inductance among devices, leading to more uniform qubits.\cite{Zhao2020, Mamin2021}
In the transmon regime where E$_J$ $\gg$ E$_C$, the plasma frequency of the junction is approximately equivalent to the qubit transition frequency (f$_{01}$ = $\frac{\sqrt{8E_C E_J}-E_C}{h})$. The typical qubit frequency of transmons is in the 4-5 GHz range.\cite{Mamin2021} 
We note that we estimate our junctions to exhibit plasma frequencies $f_{p}$ in the gigahertz range, which demonstrates promise for engineering MET devices. 

The Simmons' model estimates that the tunnel barrier width is about 25~\r{A} (fig.~4e), implying that only part of the 5~nm Zr has been oxidized in this JJ. To verify the compatibility of Simmons' model, we conduct a cross-sectional STEM annular dark-field (ADF) and electron energy loss spectroscopy (STEM-EELS) studies on the Nb/Zr~(5~nm)/O$_2$ 3700~Torr~s/Nb film from which the junction was fabricated. Fig.~4f presents an ADF-STEM image of the trilayer and reveals two distinct material compositions within the tunnel barrier, suggesting partial oxidation of Zr. 
The STEM-EELS segmentation study enables quantification of the ZrO$_2$ thickness as shown in Fig.~4g. A series of EELS O-K edges were acquired across the Nb/Zr~(5~nm)/O$_2$ 3700~Torr~s/Nb layers. Moving from the bottom Nb electrode to a top Nb electrode, we observe changes in the O-K energy loss spectra (see more details in the Appendix). We find substantial oxygen content from oxidized Zr and determine that approximately 2.5~nm of oxidized Zr. This is consistent with the effective barrier width that we have extracted using Simmons' model (Fig.~4e), which confirms the accuracy of Simmon's model in describing the tunneling characteristics of the ZrO$_2$ barrier.

Next we study how the ZrO$_x$ tunnel barriers change with nominal Zr thicknesses in the range of 2 - 5~nm using Simmon's model fits to room temperature I-V measurements. Fig.~5a and 5b show the extracted tunnel barrier height and width of the junctions fabricated from Nb/Zr/O$_2$ 3700~Torr~s/Nb films as a function of Zr thickness. We find that the barrier height of the junctions are roughly independent of thickness, except a few deviations for Zr thickness of 2~nm (Fig.~5a). The two data points showing this slight deviation are associated with the smallest device dimensions, which could lead to inconsistencies in feature resolution arising from manual fabrication steps such as resist development. The barrier width increases weakly as a function of Zr thicknesses (Fig.~5b). 
We note that junctions with Zr thicknesses of 1.5~nm and below exhibit hysteretic room-temperature behavior at high bias, which indicates electromigration through pinholes, and are thus not suitable for further transport characterization.
The minor variations may be due to process reproducibility issues, which could be mitigated through more careful calibration and optimization of the fabrication and film deposition process. The consistency of the tunnel barrier with minimal variation across different metrics, such as varying Zr thicknesses and junction sizes, is important for overall performance and stability of a tunnel junction.

\section{\label{sec:level1} Low temperature JJ characterization and subgap resistance}

Next, we cool JJ devices in a dilution refrigerator with nominal base temperature below 20~mK (filtering information given in App. H).
Their critical current densities (J$_c$) and low bias subgap resistances~(R$_{sub}$) are summarized in Table 1. R$_{sub}$ is defined as the resistance in the voltage range of 0.1-0.2~mV. 
For thicker ZrO$_2$ barriers, we find R$_{sub}$ in the $M\Omega$ range, which substantially exceeds the reported number in recent work with AlO$_x$ barriers~\cite{Anferov2024}.

We now focus on a 5~$\mu$m diameter device with a barrier defined an oxidized 5~nm thick Zr layer. 
The Ambegaokar-Baratoff relation describes the relationship between Josephson critical current (I$_c$), R$_n$, and the superconducting gap ($\Delta$) in superconductor-insulator-superconductor (SIS) systems at low temperature.\cite{Ambegaokar1963} Using R$_n$=18.98~k$\Omega$ and $2\Delta_{Nb}$=2.89~meV,\cite{Anferov2024} we estimate J$_{c,A-B}$ = 6~nA/$\mu$m$^2$, which is close to the experimental value of J$_c$= 8.14~nA/$\mu$m$^2$. 
This correspondence is consistent with the device behaving as a tunneling junction.
However, an ideal SIS system should only exhibit quasi-particle tunneling at voltages larger than 2$\Delta$. 
Our junction exhibits finite resistance at a lower voltage. 
Fig.~6c shows the numerically computed dI/dV of the I-V curve in Fig.~6a, which indicates that quasi-particle tunneling begins around 1~mV -- smaller than $\Delta_{Nb}$ extracted from the T$_c$ of Nb. 

The finite subgap resistance at intermediate voltages in our junctions might be attributable to multiple Andreev reflections (MAR),\cite{Octavio1983, Flensberg1988, Kleinsasser1994}, but our I-V curves do not exhibit the characteristic subharmonic gap structures at $2\Delta/ne$ (where $n$ is an integer and $e$ is the electron charge) in the differential conductance (dI/dV).
Indeed, attempts to fit with standard models for single-channel MAR fail to fit well, though multi-channel models could potentially account for the data. 
Future experiments that integrate these junctions to microwave devices can check for a non-sinusoidal energy-phase relation, which would imply highly transmitting channels~\cite{Willsch2024, Kim2025}.
Notably, the MAR model assumes hard superconducting gaps, meaning a sharply-defined energy gap, which may be inadequate to fully capture the subgap transport physics. 
Another possibility is the presence of strongly-coupled high-frequency modes in the environment~\cite{Hofheinz2011}, which could also be ruled out in future experiments.

\begin{table}[b]
\caption{\label{tab:table1}%
Critical current density~(J$_c$) and subgap resistance~(R$_{sub}$) of JJs with varying Zr thicknesses and diameters.
}
\begin{ruledtabular}
\begin{tabular}{c c c c} 
\textrm{Zr thickness~(nm)}&
\textrm{JJ diameter~($\mu$m)}&
\textrm{J$_c$~(nA/$\mu m^2$)}&
\textrm{R$_{sub}$~(M$\Omega$)}\\
\colrule
5 & 2 & 1.65 & 2.82\\
5 & 5 & 8.14 & 1.9\\
3 & 5 & 38.4 & 0.82\\
3 & 3 & 27.29 & 2.7\\
3 & 4 & 500 & 4.37e-3\\
1.5 & 5 & 7.31 & 7.72e-2\\
\end{tabular}
\end{ruledtabular}
\end{table}




\begin{figure*}
\includegraphics[width=\textwidth]{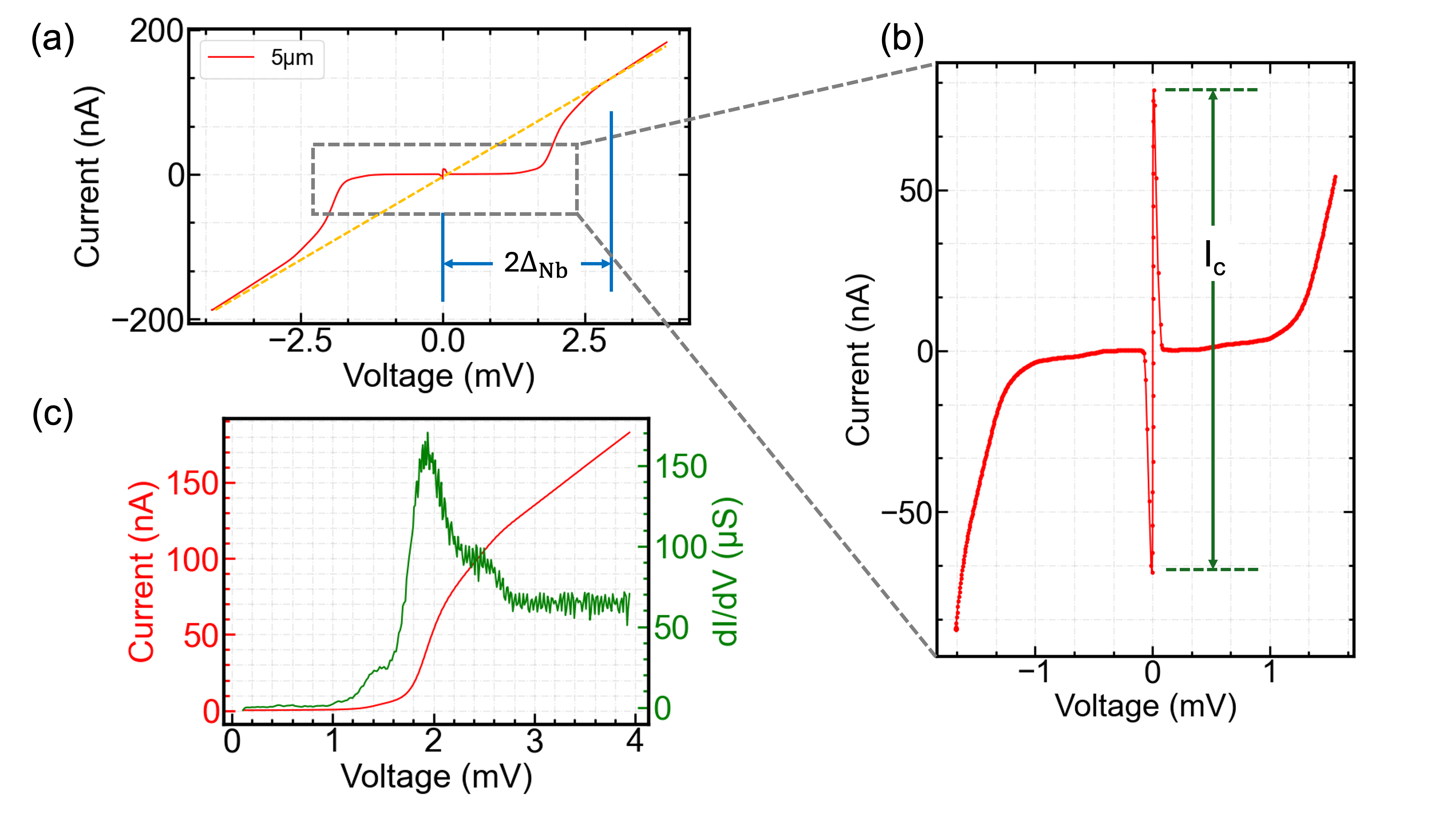}
\caption{\label{fig:wide} (a)~I-V plot of a junction grown with 5~nm of Zr at 300~mK. The normal state resistance is about 19~k$\Omega$. (b)~Magnified I-V curve showing an I$_c$ of approximately 160~nA. (c)~Numerical dI/dV of the I-V curve in (a). dI/dV shows peaks at $\pm$1.94~mV where quasi-particle tunneling occurs. This suggests that the superconducting gap is much smaller than that of pure niobium ($\Delta_{Nb}$=1.4~meV).}
\end{figure*}

\section{\label{sec:level1} Conclusion}
In this work, based on a theoretical screening of candidate materials, we identify Zr as likely to exhibit excellent oxygen retention properties. Consequently, we study ZrO$_2$ as an alternative tunnel barrier material in Nb-based JJs fabricated using a top-down, subtractive process. We then demonstrate negligible NbO$_x$ formation at the interface between the Nb electrode and the ZrO$_2$ tunnel barrier during oxidation. Simmons' model fitting allows us to extract the tunneling parameters at room temperature. We find that the extracted tunnel barrier width is consistent with the STEM-EELS analysis, indicating the compatibility of Simmons' model with our Zr-based JJs. Additionally, we observe a low-tunnel barrier height of 0.3~eV. This is potentially promising for merged-element transmon applications that benefit from a thick tunnel barrier with an I$_c$ sufficient to achieve the transmon operational impedance at lithographically easy junction areas. The low barrier height and non-uniformity of the films could contribute to the observed subgap conductance. However, the underlying mechanism and how they evolve with the thickness and composition of the Zr-based tunnel barrier remains unclear and will be investigated in a follow-up study. 

\section{\label{sec:level1}Acknowledgments}
This work was primarily supported by AFOSR (FA9550-23-1-0706). 
This work was performed in part at the Cornell NanoScale Facility, a member of the National Nanotechnology Coordinated Infrastructure (NNCI), which is supported by the National Science Foundation (Grant NNCI-2025233).

This work made use of the electron microscopy facility of the Platform for the Accelerated Realization, Analysis, and Discovery of Interface Materials (PARADIM), which is supported by the National Science Foundation under Cooperative Agreement No. DMR-2039380 and Cornell Center for Materials Research (CCMR) shared instrumentation facility with Helios FIB supported by NSF (DMR-1539918), FEI Spectra 300 acquired through NSF-MRI-1429155 and for XPS surface analysis. The authors also thank John Grazul, Mariena Silvestry Ramos and Malcolm Thomas for technical support and maintenance of the electron microscopy facilities.
This work made use of the Meehl cryostat donated by David W. Meehl in memory of his father James R. Meehl and supported by the Cornell College of Engineering.
We also thank Joël Griesmar, Jean-Damien Pillet, and Landry Bretheau for discussions and feedback on interpreting the low-temperature I-V data. 

\section{\label{sec:level1}Author Declarations}
Cornell University (D.A.M.) has licensed the EMPAD hardware to Thermo Fisher Scientific. The remaining authors declare no competing interests.

\section{Author Contributions}
\textbf{Jaehong Choi}: Conceptualization (equal); Formal analysis (lead); Investigation (lead); Visualization (equal); Methodology (equal) Writing/original draft preparation (lead).
\textbf{Maciej Olszewski}: Investigation (equal); Methodology (equal).
\textbf{Luojia Zhang}: Investigation (equal); Visualization (equal); Methodology (equal).
\textbf{Zhaslan Baraissov}: Investigation (equal); Methodology (equal); Formal analysis (equal); Visualization (equal); Writing/review \& editing (supporting). 
\textbf{Tathagata Banerjee}: Investigation (equal); Methodology (equal); Formal analysis (equal); Visualization (equal).
\textbf{Kushagra Aggarwal}: Investigation (supporting); Methodology (supporting).
\textbf{Sarvesh Chaudhari}: Investigation (equal); Methodology (equal); Formal analysis (equal). 
\textbf{Tom\'as A. Arias}: Investigation (equal); Methodology (equal); Supervision (equal); Writing/review \& editing (supporting).
\textbf{David A. Muller}: Methodology (equal); Supervision (equal); Writing/review \& editing (supporting).
\textbf{Valla Fatemi}: Conceptualization (equal); Writing/review \& editing (supporting). 
\textbf{Gregory D. Fuchs}: Conceptualization (equal); Investigation (lead); Methodology (equal); Writing/review \& editing (lead).

\section{Data Availability}
The data that support the findings of this study are openly available in [insert Cornell eCommons URL on publication].

\section{Appendices}
\appendix

\section{Density functional study of the tunnel barrier materials}
We use the open source cope Joint Density Functional Theory (JDFTx) and the $\Delta$E calculations are done using plane wave bases and periodic boundary conditions within the generalized gradient approximation (GGA). All defect and bulk calculations for the materials are conducted using their supercells, and we utilize the same supercell for the bulk and defect structure of a given material for consistency.

\section{STEM sample preparation}
The TEM lamella used in this study were prepared with ThermoFisher Helios Focused ion beam using standard lift-out method. The milling of the lamella was done with voltages starting from 30 kV down to 2kV during the final milling step.  HAADF STEM images in Fig.~1a-c and Fig.~2b are acquired with aberration-corrected ThermoFischer Spectra 300 at 300 keV beam energy and 30 mrad semiconvergence angle. The Nano-beam electron diffraction experiments were done using the FEI TitanThemis at 300 kV and convergence semiangle of 1 mrad. The 4D-STEM datasets were acquired with EMPAD-G2 detector.\cite{Phillipp2022}

\section{EELS acquisition and segmentation}

\begin{figure*}[h!]
\includegraphics[width=\textwidth]{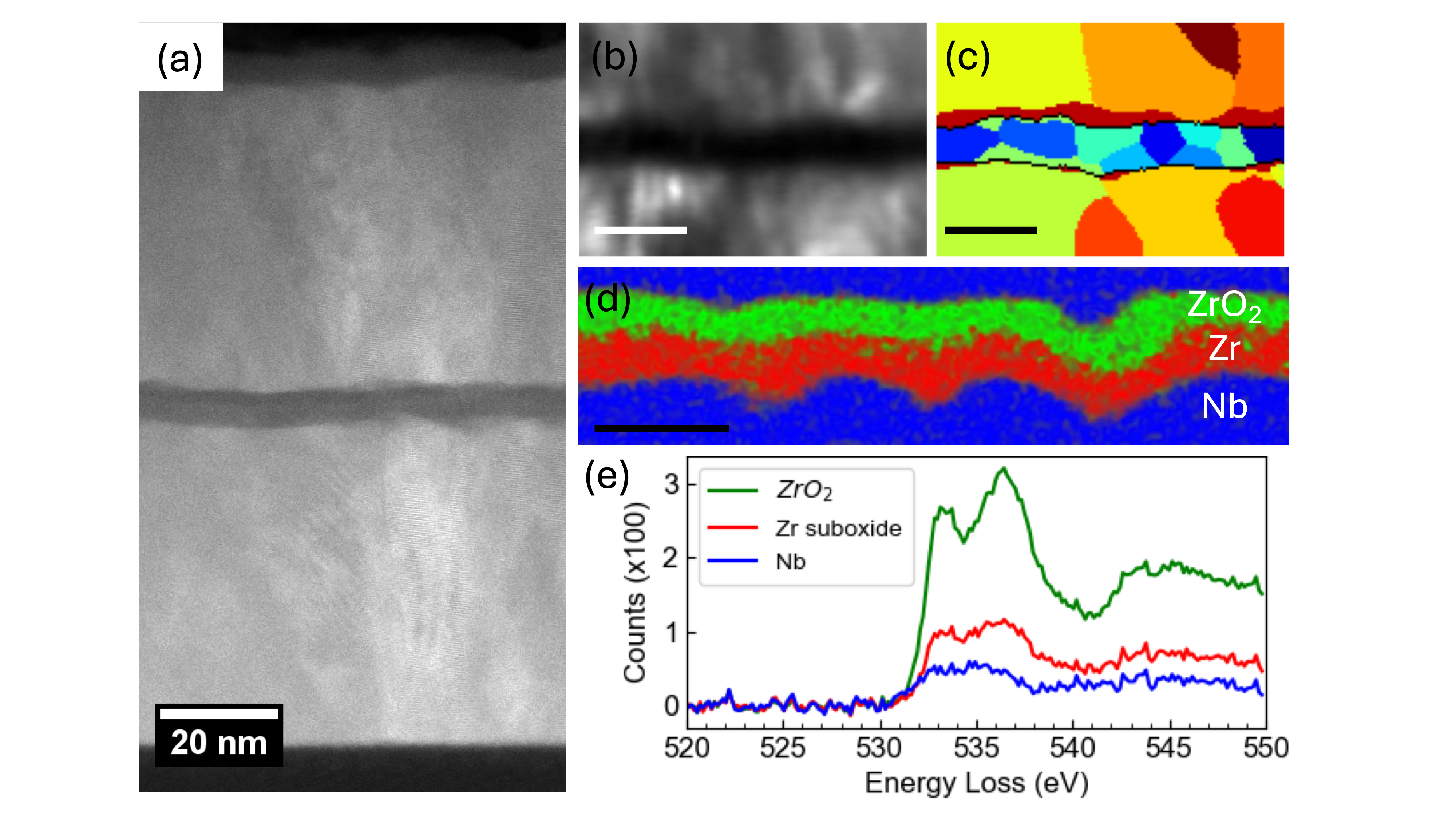}
\caption{\label{fig:wide} EELS acquisition and segmentation. (a)~Low magnification HAADF–STEM image showing the cross-section of the device. The substrate on the bottom is Si, with 5 nm of ZrO$_2$ sandwiched in between two 50~nm thick Nb layers. The top Nb layer is exposed to air and has formed the native Nb oxide.  (b)~Virtual ADF image obtained from Nano-Beam Electron Diffraction (NBED) data, collected with 1 mrad convergence semiangle. (c)~Segmented real-space image obtained from NBED data after cepstral transform and k-means clustering. The ZrO$_2$ grain size ranges from 3-5~nm, while Nb grains exhibit columnar structure with lateral size of 10-15~nm. The colors do not represent orientation. Dark red region on the interface represents area with the overlap of both Nb and ZrO$_2$ diffraction patterns. (d)~Segmented EELS image, showing separation into three distinct regions. The Nb electrodes are shown in blue, while the tunneling barrier has the fully oxidized layer ZrO$_2$, shown in green, and the sub-oxide region with disordered oxygen shown in red. (e)~Three O-K spectra averaged over the regions shown in (d). The blue spectrum corresponds to the signal coming from the Nb oxide formed on the side surfaces of the lamella and interstitial oxygen in Nb. Similarly, the signal from the red region represents the surface Zr oxide and potential sub-oxide formed during oxidation of Zr. The green spectrum corresponds to the fully-oxidized tetragonal ZrO$_2$ oxide. The size of the scalebars in (b)-(d) is 10 nm.
}
\end{figure*}

The EELS data was acquired using aberration-corrected ThermoFischer Spectra 300 with Continuum CMOS camera. We used an electron beam with energy of 120 keV and current of 150 pA, convergence semiangle of 30 mrad and EELS collection angle of 100 mrad. The energy dispersion was chosen at 0.15 eV, while the FWHM of zero-loss peak is estimated at 0.6 eV. The field of view is 52.5 $\times$ 13.1 nm with about 1 Å scanning step size. The spectra were not deconvolved, the average thickness in the field of view is ~0.4 inelastic mean free path.
For the segmentation part, the EELS datasets were first background-subtracted around the Nb-M3,2, Zr-M3,2, and O-K edges. Next, these three regions of the spectrum were stitched together and subjected to dimensionality reduction using diffusion mapping.\cite{Colletta2023} The resulting basis was then separated into three different clusters using the fuzzy c-means clustering algorithm.


\section{Wafer preparation}
All depositions are performed on 525 $\mu$m thick 100~mm float-zone silicon (100) wafers with resistivity $\geq$ 10 k$\Omega$ cm (WaferPro). Prior to deposition a two-step RCA clean was performed: SC-1 and SC-2. SC-1 is a 10-min soak in a 6:1:1 mixture of DI water, ammonium hydroxide, and hydrogen peroxide, respectively. The clean is done at 70$^{\circ}$C, followed by a 10 minutes rinse in DI water. SC-2 is a 10 minute soak at 70$^{\circ}$C in a 6:1:1 mixture of DI water, hydrochloric acid, and hydrogen peroxide, respectively. The clean is followed by a 10 minute DI water rinse and nitrogen gas dry. Within one hour of completing the RCA clean the silicon oxide is removed using a 10:1 buffered oxide etch (BOE) soak for 1 minute, two 30-second deionized water rinses, and a blow dry with nitrogen. The wafer is loaded into the deposition chamber load lock within 5 min of concluding the BOE etch process. We pump on the load lock for 1.5 to 2 hours, reaching pressures below $8\times10^{-7}$ Torr before transferring the wafer into the chamber for deposition.

\section{quadrilayer deposition}
The JJ film depositions were done in the spintronics deposition chamber discussed in Olszewski et al.\cite{Olszewski2025} The first 60~nm Nb deposition was done at 2~mTorr and 70~sccm flow of ultra high-purity Ar gas from Airgas and a 3N5 purity, 2 inch diamter, and 1/8~inch thick target from AJA international Inc. With a target-to-substrate distance of around 20 to 30~cm, these depositions had a rate of about 2.4~\r{A}/s at 325~W. The substrate was rotated at 50~rpm during the depositions. The base pressure varied between $1\times10^{-9}$ to $2\times10^{-9}$~Torr. 

Following the Nb deposition, the Zr of varying thickness was deposited with the same process gas parameters using a 3N5 purity, 2 inch diameter, and 1/8 inch Zr thick target from AJA International Inc. The rate was about 2.0~\r{A}/s at 200~W.
Upon completion of the Nb and Zr deposition, the chamber was purged and subsequently flooded with a 15\% oxygen in argon mixture with a 100~sccm flow rate. After 16~min of continuous gas flow, the pressure reaches around 4.2~Torr, at which point the oxidation gas is purged. The chamber is then pumped for around 2 to 3 hours to remove excess oxidation gas. Before depositing the top Nb layer, the Nb target is pre-sputtered for 5 minutes, before depositing the top 60~nm layer of Nb with the same deposition parameters as the bottom Nb layer.

\section{Fabrication details}
We sonicate the trilayer in acetone and isopropyl alcohol (IPA) for 5~minutes each. Then, we spin coat the wafer with P20 primer and S1813 photoresist, followed by a pre-bake at 115~$^{\circ}$C for 1 minute. The wafer is then exposed to a 436~nm laser in the DSW 5X g-line wafer stepper system. After exposure, we develop the wafer in AZ MIF 726 for 45 seconds, followed by a DI water rinse. Next, the wafer is pre-etched in oxygen plasma to remove photoresist residue and then etched using a chlorine based plasma (Cl$_2$, BCl$_2$, Ar) in the PT770 ICP-RIE etcher for bottom electrode characterization. We do not etch the trilayer all the way down to the the silicon substrate during this step because the subsequent etching step for tunnel barrier definition will inevitably increase the bottom Nb electrode height by further etching into the silicon. This reduces the conformity of the photoresist to the structure and can lead to unwanted SiO$_2$ scaffold over-etching in the buffered oxide etch (BOE) process later, potentially causing shorts between the top and bottom electrodes. After the partial etching, we strip the resist in AZ 300T at 80~$^{\circ}$C for 10 minutes, sonicate the sample in IPA for 10 minutes, and rinse with DI water. Then, we dehydrate the wafer at 180~$^{\circ}$C for 3 minutes and spin coat LOR 3A , followed by a pre-bake at 180~$^{\circ}$C for 4 minutes for tunnel barrier patterning. Next, we spin coat S1813 and pre-bake it at 115~$^{\circ}$C for 1 minute, and expose the wafer again in the DSW 5X g-line wafer stepper. The wafer is developed in AZ MIF 726 for 30 seconds, followed by a DI water rinse. After oxygen plasma pre-etching, the tunnel barrier is defined in chlorine-based plasma. Without removing the resist, we load the wafer into the evaporator and deposit approximately 100~nm of SiO$_2$ as a scaffold layer. After deposition, the wafer is immersed in AZ 300T stripper overnight for lift-off. Following lift-off, we perform another hot AZ 300T bath at 80~$^{\circ}$C for 10 minutes, followed by IPA sonication and DI water cleaning. At this point, the entire wafer is covered with SiO$_2$. To selectively etch SiO$_2$ and leave only a small area around the tunnel barrier, we spin coat and expose another layer of S1813 resist. After development in AZ MIF 726, we etch the SiO$_2$ in BOE 10:1 and rinse with DI water. After BOE etching, square-shaped SiO$_2$ with a 3~$\mu$m overhang remains around the tunnel barrier. We then remove the photoresist in acetone and IPA and then perform the final round of resist coating of LOR5A and S1813, each followed by baking at 180~$^{\circ}$C and 115~$^{\circ}$C for 4 minutes and 1 minute, respectively. After exposure in the 5X stepper and development, the wafer is loaded into the AJA sputter system for top electrode deposition, following oxygen plasma pre-etching. We perform Ar etching to remove the thin NbO$_x$ layer formed on the top electrode surface and sputter roughly 300~nm of Nb as the top electrode without breaking vacuum. After sputtering, the wafer is left in 1165 overnight, followed by a hot 1165 bath at 80~$^{\circ}$C for 10 minutes, IPA sonication, and DI water rinse. Finally, the wafer is annealed at 180~$^{\circ}$C for dehydration, and the SiO$_2$ scaffold is etched using vapor hydrofluoric (HF). Each chip has eight junctions with varying diameters from 2~$\mu$m to 9~$\mu$m.

\section{XPS analysis of the Nb/Zr-ZrO$_x$/Nb trilayer}
XPS data is collected in a Thermo Nexsa G2 Surface Analysis System with a chamber pressure of around $3\times10^{-7}$ Torr. The Nexsa uses a monochromated Al K alpha x-ray source with the fermi level calibrated with a silver standard. Survey scans are collected as counts per second (CPS) with a 0.4~eV energy resolution using an x-ray spot size of 400~$\mu$m. Etching is done using a monoatomic Ar ion gun with an ion beam spot size of 1~mm, a voltage of 2000~eV and low beam current. Data was collected after every 60~s of etching. Core level spectra are collected with a 0.1~eV energy resolution, and fitting is done in CasaXPS (v2.3.25)\cite{Fairley2021} with a Shirley background using Thermo Scientific provided sensitivity factors calibrated for the Nexsa. Peaks are fit with asymmetric Voigt-like lineshapes for the Nb and Zr metals, and symmetric gaussian-lorentzian mixed lineshapes for the oxides. 

The Nb signal at the surface of the sample is composed of metallic Nb (8.3\%), NbO (3.1\%) NbO\textsubscript{2} (7.8\%) and primarily Nb\textsubscript{2}O\textsubscript{5} (80.8\%).

The Zr signal seen in Fig. 1(f) has about 68.7\% Zr, 24.3\% ZrO\textsubscript{2}, and 7\% suboxide.
The Zr oxide signal is a combination of both non-stoichiometric oxide formed during oxidation and damage from the destructive Ar ion milling process. 
Unfortunately, it is difficult to identify the composition of the suboxides using XPS analysis. 
As we move from the top Nb electrode across the tunnel barrier layer, we observe ZrO$_x$ and Zr metal in the barrier layer, confirming that Zr is not fully oxidized when we deposit 5~nm of Zr. 
The profile away from the ZrO$_x$ layer is identical to that of pure Nb films, thus indicating that there is no additional oxide formation or leakage from the tunnel barrier into the electrodes.
We note that the XPS data alone cannot completely rule out the formation of NbO$_x$ due to limitations in deconvoluting a potential weak NbO$_x$ signal during peak fitting possibly caused by the close proximity of peaks and the dominant Nb metal signal. 

\section{Low temperature I-V characteristics of JJs with different Zr thicknesses} \label{Sec:low-T-IV}
We explore junctions with ZrO$_2$ thicknesses other than 5~nm. Fig.~8a–e show the low temperature characteristics of JJs with 3~nm and 1.5~nm thick ZrO$_2$ barriers, while figure 9 shows the measurement circuit diagram. The JJ with a 3~nm thick ZrO$_2$ layer exhibits an I$_c$ of 1.7~$\mu$A and a normal resistance of 1.589~k$\Omega$ (Fig.~8a–b). Its subgap resistance is 150~k$\Omega$, and it displays hysteresis (Fig.~8c). For the JJ with a 1.5~nm thick ZrO$_2$ barrier, I$_c$ is 45.7~nA, and the normal resistance is 2.37~k$\Omega$ (Fig.~8d–e). The subgap current is noisy, and the subgap resistance is smaller—77.2~k$\Omega$—compared to those of the 3~nm and 5~nm JJs. This may indicate low tunnel barrier quality, such as non-uniform thickness. Since the film growth process has not yet been optimized, further improvement in film quality is expected after optimization. Additionally, the high subgap conductance observed in the JJ with a thinner barrier suggests that the subgap current is not due to a proximity effect.
\begin{figure*}[h!]
\includegraphics[width=\textwidth]{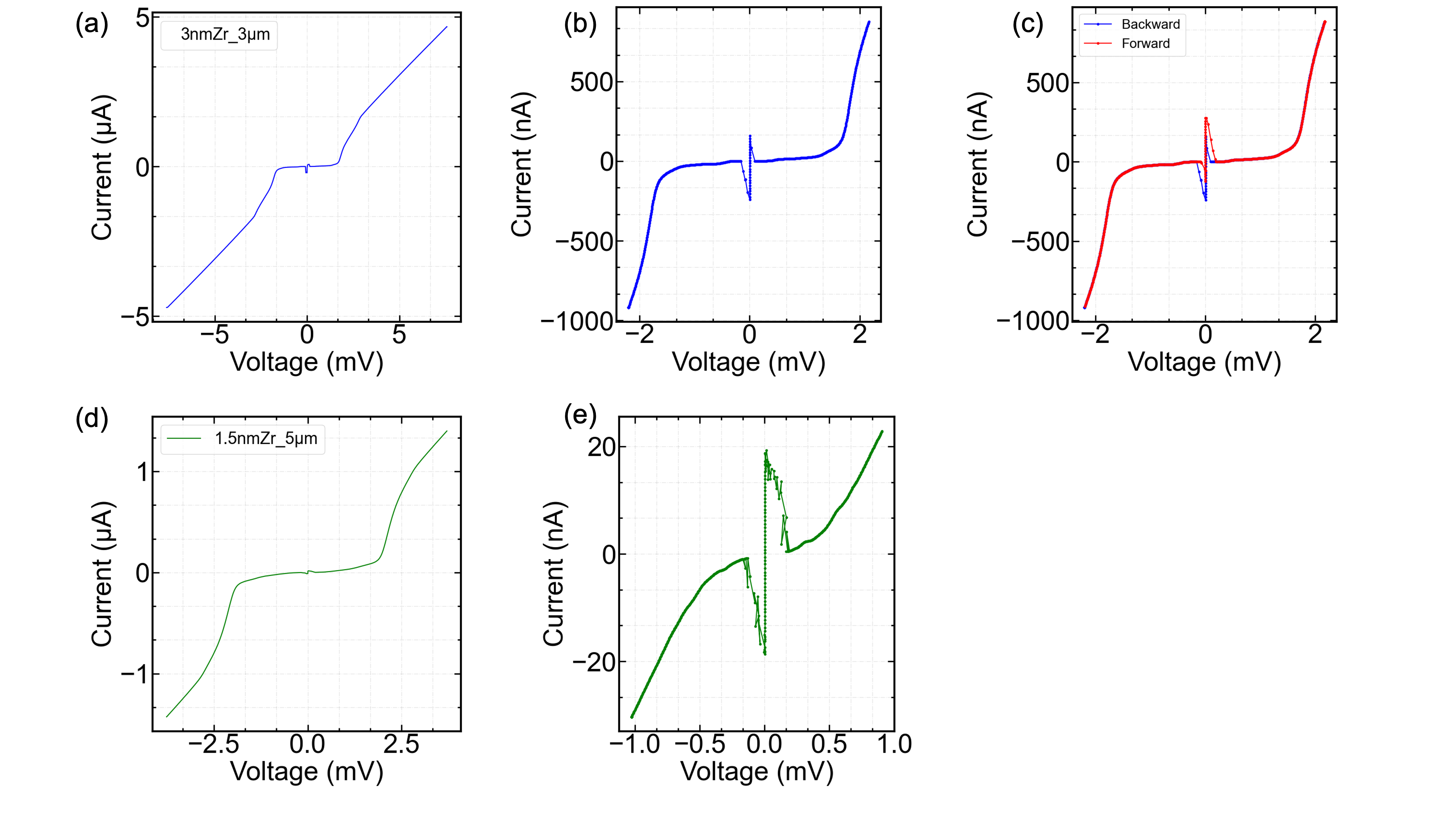}
\caption{\label{fig:wide} Low-temperature I-V characteristics of junctions with nominal Zr thicknesses of 3~nm and 1.5~nm.
(a) I-V curve of a junction with nominal 3~nm Zr, and (b) magnified I-V of this junction. (c) The 3~nm Zr junction shows a hysteretic I-V, indicating high junction quality. (d) I-V curve of the junction with nominal 1.5~nm Zr. (e) Magnified view of the I-V. The noisy I-V may suggest a low-quality tunnel barrier. Optimization of the film deposition process is needed to achieve uniform thin films. This junction did not show a hysteretic I-V. 
}
\end{figure*}
\begin{figure*}[h!]
\includegraphics[width=\textwidth]{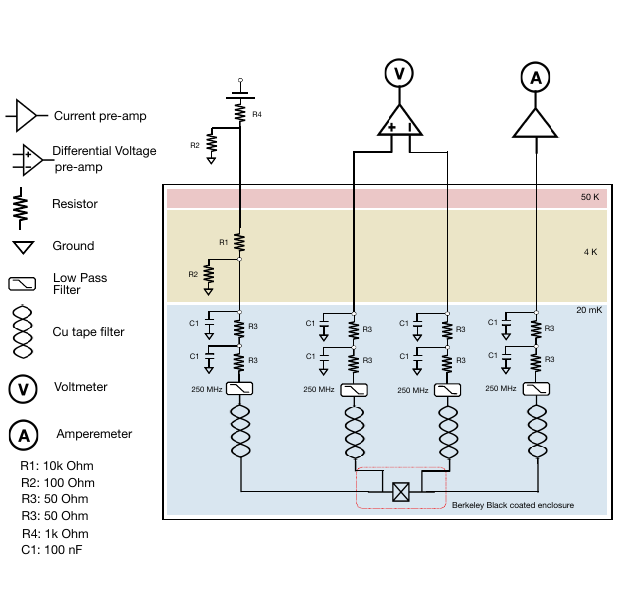}
\caption{\label{fig:wide} Circuit diagram for measuring low temperature Josephson I-V curves.
}
\end{figure*}

\section{Nb/Al/AlO$_x$/Al/Nb Josephson junction}
We sputter 50~nm of Nb and 10~nm of Al onto a chemically etched silicon substrate. Then, oxygen is introduced to the chamber at 100~Torr for one hour. Afterward, another 10~nm of Al and a 50~nm of top Nb are deposited. We fabricate the JJs using the same fabrication processes as for the ZrO$_2$-based JJs. For a JJ with a 2~$\mu$m diameter, the Simmons' model estimates the tunnel barrier height to be 2.8~eV and the width to be 1~nm.

\clearpage
\nocite{*}
\bibliography{aipsamp}

\end{document}